\documentclass[%
 reprint,
 amsmath,amssymb,
]{revtex4-1}

\usepackage{dcolumn}% Align table columns on decimal point
\usepackage{bm}% bold math
\usepackage{color}
\usepackage{graphicx,amsmath,amssymb,xspace,epsfig,float,multirow,subfigure,tabularx}
\begin{document}

\preprint{APS/123-QED}

\title{Nonlinear transport by vortex tangles in cuprate
high-temperature superconductors}
% Force line breaks with \\

\author{Rong Li}

\author{Zhen-Su She}%
 \email{she@pku.edu.cn}
\affiliation{%
State Key Laboratory for Turbulence and Complex Systems,  College of Engineering, Peking University, Beijing 100871, China}%

\date{\today}

\begin{abstract}
A unified model of vortex tangles is proposed to describe unconventional transport in cuprate high-temperature superconductors,
which not only captures the fast vortices scenario at low density, but also predicts a novel mechanism of core-core collisions in dense vortex fluid regime. The theory clarifies the nature of vortex fluctuations being the quantum fluctuations of holes and then resolves
a discrepancy of two orders of magnitude of Anderson's damping model $\hbar n_v$, with right prediction of the nonlinear field dependence of the resistivity $\rho=\rho_n(B+B_T)/(B_0+B+B_T)$ and the Nernst effect, validated by data of several samples. Consequently, Anderson's vortex tangles concept and phase fluctuation scenario of pseudogap are verified quantitatively.
\end{abstract}

\pacs{Valid PACS appear here}% PACS, the Physics and Astronomy
%Use showkeys class option if keyword
\maketitle

%\section{Introduction}\label{intro}

{Recent experimental discoveries of the weak diamagnetism \cite{Li2010} and strong Nernst signal \cite{Wang2006} above $T_c$ have
stimulated a hot debate about the presence of vortex liquid in pseudogap regime in high-temperature superconductivity (HTSC).
A theoretically proposed fast-vortex scenario \cite{Ioffe2002} is experimentally found \cite{Bilbro2011} in dilute vortex regime, which is followed by a quick saturation in the magnetoresistance at high fields (¡°knee¡± feature) {below $T_c$} \cite{Wang2006}, or by a weak field dependence {at high temperature in pseudogap state} \cite{Usui2014,Lee2006}. The {saturation} is an unconventional behavior associated with dense vortex fluid, unable to be explained by isolated vortex scenario, nor by
Bardeen-Stephen model \cite{Bardeen1965} as well as the fast vortex theory \cite{Ioffe2002}.
In the dense vortex fluid, quasiparticle (qp) density of state (DOS) can be remarkably influenced by the overlapping of qp wave functions of neighboring vortices \cite{Melnikov2006,Canel1965}, leading to an enhancement of qp scattering due to vortex-vortex
interaction. This possibility was explored by Anderson with the idea of vortex tangles \cite{Anderson0607}, but his estimated
damping coefficient $\eta=\hbar n_v$ due to quantum fluctuations \cite{Anderson06} presents an overestimation of two orders of
magnitude compared to data. Moreover, current vortex fluid models and simulations can only account for the transverse thermoelectric coefficient $\alpha=e_N/\rho$ or Nernst signal, qualitatively. Thus, the legitimacy of the vortex liquid scenario and, more importantly the phase fluctuation explanation of pseudogap, calls for a refined quantitative damping model of vortex tangles.

In this letter, Anderson's idea of vortex tangles is extended to form a unified model for dilute and dense, magnetic
and thermal vortex fluid. A novel damping mechanism of vortex tangles is proposed, which describes both the qp-defect
scattering from isolated vortex and a novel core-core collisions of vortex entanglement. The new model clarifies the nature of the vortex fluid being the quantum fluctuations of holes, which resolves the discrepancy of Anderson's model, and describes quantitatively both the nonlinear field dependence of flux-flow resistivity $\rho=\rho_n(B+B_T)/(B_0+B+B_T)$ and the Nernst effect, as validated by data of several cuprate HTSC samples. Thus, a unified description of vortex transport in HTSC is achieved, establishing the legitimacy of Anderson's vortex tangles concept and phase fluctuation explanation of pseudogap \cite{Emery1995}.}

%\section{Damping viscosity of vortex tangles}

We begin with a well-known picture of vortex fluid composed of pancake vortices, when a magnetic field is applied perpendicular to Cu-O plane  on a cuprate superconductor. Once the temperature is higher than the critical
temperature of Berezinskii-Kosterlitz-Thouless (BKT) phase transition \cite{BKT}, thermal vortices unbind, thus the vortex density $n_v=(B+B_T)/\phi_0$, where $B_T$ is the characteristic field defined to describe the density of
thermal vortices.  When a vortex moves at a velocity $v$, the damping force is $f_{d}=\eta v$, where
$\eta$ is the damping coefficient. The resistivity due to damping transport of magnetic and thermal vortices is \cite{Halperin1979}
\begin{eqnarray}
\rho={n_v\over\eta}\phi_0^2c_0.
\label{rho}
\end{eqnarray}
 By the dimensional analysis,
\begin{eqnarray}
\eta={m_p\over \tau},
\end{eqnarray}
where {$m_p$ is the effective mass of the core} and $\tau$ is the characteristic damping time. Both quantities will be estimated below based on our entangled vortex fluid model.

In a entangled vortex fluid, random motions of vortices are affected by three types of fluctuations: thermal fluctuations,
quantum fluctuations of vortices, and quantum fluctuations of holes inside vortex cores. Their relative importance can be
determined by the following energy estimates. The energy of quantum-fluctuations of vortices is $\epsilon_v=\hbar^2n_v/2m_p$,
where the vortex effective mass $m_p$ is defined as the bare hole mass inside the vortex core: $m_p=\pi\xi^2n_hm_e$, with $\xi$
the coherence length, $n_h$ the two-dimensional hole density on Cu-O plane (multiply by layers), and $m_e$ the electron mass.
On the other hand, the energy associated with quantum fluctuations of holes is $\epsilon_h=\hbar^2n_h/2m_e$. Below, we express
the three characteristic energy in terms of critical temperature $T_c$, critical vortex density $n_{c2}=H_{c2}/\phi_0$ and hole
density; for instance, for optimal doped (OP) Bi$_{2}$Sr$_{2}$CaCu$_{2}$O$_{8+\delta}$ (Bi-2212) \cite{Wang2003,Zhou1999},
$\epsilon_v=0.84 k_B\ll\epsilon_T=90k_B\ll\epsilon_h=478k_B$ (in Joule unit). This indicates that the fluctuations are dominated
by quantum fluctuations of holes inside vortex cores. Using $\pi\epsilon_h$ in which a geometric factor $\pi$ is considered, the
random velocity is then found to be
\begin{eqnarray}
v_R={\hbar\over{m_e\xi}}.
\label{vR}
\end{eqnarray}

\begin{figure}
  \centering
  \includegraphics[width=8cm]{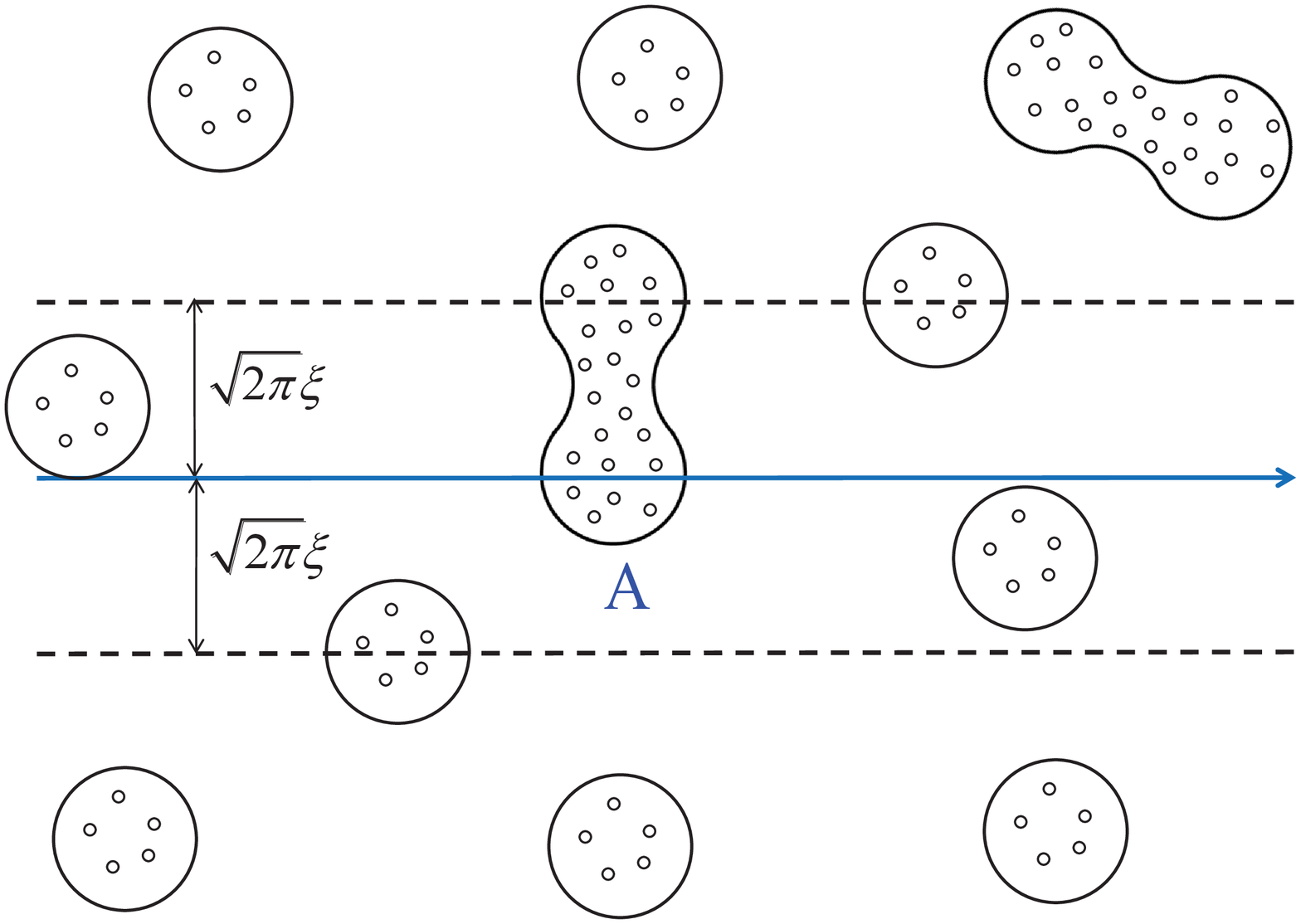}\\
  \caption{Schematic diagram of core-core collisions. Vortex cores and qp are represented by big and small black
  circles, respectively. The qp DOS inside the isolate vortex core are small, thus should be described
  by a fast vortex scenario. On the other hand, in a core-core collision process, two vortex cores merge together to
  a bigger core which leads to a increase of qp DOS and damping. Two dash lines indicate that only the
   vortices the distance between which and the core trajectory (blue and solid line) is less than $\sqrt{2\pi}\xi$
   participate in core-core collisions with vortex A.}
  \label{CCColl}
\end{figure}

We propose that this speed controls the core-core collisions. Fig. \ref{CCColl} schematically shows the process of core-core
collision: two vortex cores approach each other and merge into a bigger core due to qp wave-functions overlapping, leading to a
decrease of intervals between the qp energy levels and the increase of low-energy DOS, as well as the enhancement of the
qp-defect scattering inside the bigger core. Subsequently, the bigger core decompose into two cores which separate from each
other. The critical vortex distance at which qp wave functions of neighboring vortices merge together is $\sqrt{2\pi}\xi$, i.e.,
at $B=H_{c2}$, but not the penetration depth $\lambda$ according to the modified London equation \cite{Tinkham1996}. The reason
is that when the vortex distance $l_v$ is much larger than $\sqrt{2\pi}\xi$, the influence of one vortex on the other do not
change the qp wave functions inside vortex cores, and so do not contribute to dissipation. Thus, we define $\sqrt{2\pi}\xi$ as
the core-core scattering length, below which the core-core collision takes place. This yields an estimate of the mean-free-path
for a core-core collision as
\begin{eqnarray}
l_{c}=1/2\sqrt{2\pi}n_v\xi,
\label{lc}
\end{eqnarray}
where we assume that the dissipation is similar for thermal and magnetic vortex. The characteristic time of core-core collision
is then
\begin{eqnarray}
{\tau_{c}={l_{c}\over v_R}={m_e\over{2\sqrt{2\pi}\hbar n_v}}.}
\label{tauc}
\end{eqnarray}

The momentum loss of two colliding vortices in one core-core collision determines the damping force, thus the damping coefficient $\eta_c$. Cuprate superconductors are doped Mott insulators in which plenty of defects (e.g., oxygen vacancies) exist on Cu-O plane \cite{Blatter1994}, {thus the relaxation time of qp scattering $\tau_q=m_ec_0/n_he^2\rho_n$ is less than the core-core collision time $\tau_c$ in most regimes of vortex fluid phase except near $H_{c2}$, where $c_0$ is the lattice constant of c axis. For example, in OP Bi-2212, $\tau_q=7.6\times10^{-14}$ s, and $\tau_c=B^{-1}3.6\times10^{-12}$ s, indicating $\tau_q\leq\tau_c$ at $B\leq47$ T.} Therefore, it is reasonable to assume the transport momentum of two vortices totally disappear after one collision. This yields $\eta_c=m_p/\tau_c$. Eq. (\ref{vR}) and (\ref{tauc}) yields
\begin{eqnarray}
\eta_c=(2\pi)^{3/2}n_h\xi^2\hbar n_v.
\label{etac}
\end{eqnarray}

{A comparison between Eq. (\ref{etac}) and Anderson's original model reveals the true physics of
vortex tangles.  Anderson proposed a vortex-vortex collision picture driven by quantum fluctuations of vortices
(i. e. $\epsilon_v$), predicting a damping coefficient $\eta=\hbar n_v$ with a random velocity $\hbar/m_pl_v$
and mean-free-path $l_v$ \cite{Anderson06}. Substitute it into the resistivity formula Eq. (\ref{rho}), this yields
$\rho=\rho_n=\hbar\phi_0^2c_0$, independent of magnetic field, temperature and doping. A huge deficit arises since the
predicted $\rho_n$ (e. g., 126 $\mu\Omega$m for any doping of Bi-2212) is nearly two orders of magnitude higher
than experimental data (of Bi-2212) \cite{Usui2014}, see the red suqares in Fig. \ref{rhon}.
This discrepancy is now resolved in Eq.(\ref{etac}): a numerical factor $(2\pi)^{3/2}n_h\xi^2$ (equals 82 for OP Bi-2212) \cite{Wang2006} is recovered due to the fact that the damping mechanism of vortex tangles is not driven by quantum fluctuations of vortices (as a whole) but by quantum fluctuations of holes inside the vortex cores.}

%\section{Nonlinear flux-flow resistivity}
 At the dense vortices limit, core-core collisions dominate the damping coefficient, and $\rho$ approaches the normal state resistivity,
\begin{eqnarray}
\rho_n=\sqrt{\pi\over2}{c_0H_{c2}\over{en_h}}.
\label{rhonpre}
\end{eqnarray}
{Qualitatively, $H_{c2}$ shows a weak $T$ dependence below and near $T_c$ in HTSC \cite{Wang2006},
thus Eq. (\ref{rhonpre}) predicts a constant $\rho_n$ in this regime, which is consistent with
data \cite{Zhang2000,Ando1996}.}
Fig.\ref{rhon} shows the comparisons between theoretical predictions of Eq. (\ref{rhonpre}) (taking $H_{c2}$
reported in \cite{Wang2003,Li2010,Ando1996})
and the experimental data of $\rho_n$ at the onset temperature at different doping \cite{Usui2014,Ri1994,Zhang2000,
Ando1996}. In making the prediction, the hole density $n_h$ is set by $n_h=n*p*(a_0*b_0)^{-1}$ with the layer number
$n$, lattice constants $a_0$ and $b_0$ of the Cu-O plane \cite{Zhou1999}, and hole concentration $p$ which is
estimated from the empirical formula ${T_c}(p) = T_{c,\max }[1 - 82.6{(p -0.16)^2}]$, where $T_{c,\max }$ is the maximum
$T_c$ in one material \cite{Obertelli1992}. The agreement is very satisfactory with errors of the same order as the experimental uncertainty ($\pm10\%$) of $\rho_n$ and $H_{c2}$; thus, {the damping model of core-core collisions Eq. (\ref{etac}) is
reliably verified. By the way, a more precise comparison requires the specification of
the temperature dependence of $H_{c2}$ and of the damping time $\tau$ (from $\tau_c$), which will
be discussed elsewhere.

\begin{figure}
\centering
\includegraphics[width=8cm]{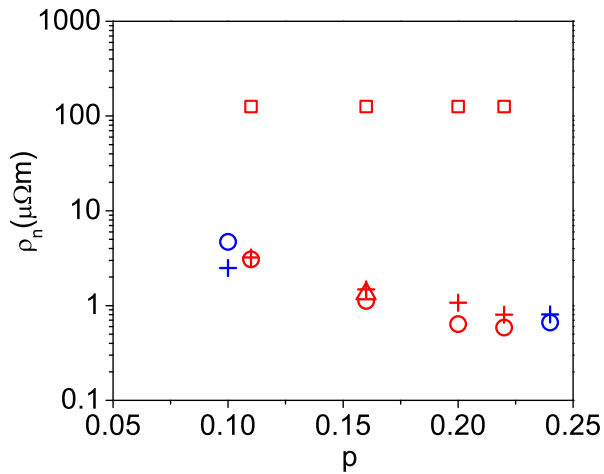}\\
\caption{$\rho_n$ vs p (hole concentration) predicted with Eq. (\ref{rhonpre}). The red and blue circles are from
the experimental measurements of Bi-2212 \cite{Usui2014}, and  Bi-2201 \cite{Zhang2000,Ando1996} respectively, while the red triangle represents another sample of OP Bi-2212 \cite{Ri1994}. The crosses are predictions of Eq. (\ref{rhonpre}) with measured $H_{c2}$: red for Bi-2212 from Nernst signal \cite{Wang2003}, blue for Bi-2201 from diamagnetism \cite{Li2010} (at p=0.10, taking the data at p=0.095 as a approximation) and from magnetoresistance \cite{Ando1996} (at p=0.24), respectively. The red squares are predictions from Anderson's damping model \cite{Anderson06}. }
\label{rhon}
\end{figure}

Now, let us make a specific model for $\eta$. Generally speaking, three sources contribute to damping force, namely qp scattering
from isolated vortex, the core-core collisions, and the pinning \cite{Li2017}. Therefore, $\eta$
can be expressed as
\begin{eqnarray}
\eta=\eta_0+\eta_{c}+\eta_{pin},
\label{eta}
\end{eqnarray}
where $\eta_0$ represents qp-defect scattering from isolated vortex and is a material parameter independent of
temperature and fields,  $\eta_c$ core-core collisions, and $\eta_{pin}$ the pinning effect which is negligible in
dense vortex fluid (e.g strong field or high temperature). Since $\eta_0$ is a constant, one
can define an effective field $B_0$ so that
\begin{eqnarray}
\eta_0=B_0\phi_0c_0/\rho_n.
\label{eta0}
\end{eqnarray}
According to Eq. (\ref{etac}) and (\ref{rhonpre}), $\eta_c=n_v\phi_0^2c_0/\rho_n$. Combining Eq. (\ref{eta}) and (\ref{eta0}) and neglecting the pinning effect, one can obtain
\begin{eqnarray}
\rho=\rho_n{{B+B_T}\over{B_0+B+B_T}}.
\label{rhoBpre}
\end{eqnarray}
Comparing to BS model \cite{Bardeen1965}, the extra terms $B+B_T$ in the denominator represent the damping effects due to core-core collisions of magnatic and thermal vortices,
thus $\rho$ nonlinearly depends on vortex density. In Fig. \ref{rhoB}, predictions of Eq. (\ref{rhoBpre})
(solid lines) are compared with the magnetoresistance data of overdoped (OD, p=0.2) La$_{2-x}$Sr$_{x}$CuO$_{4}$ (LSCO) below
$T_c$ and OP Bi-2212 in the pseudogap state \cite{Wang2006,Ri1994}. $\rho_n$ used here is predicted with
Eq. (\ref{rhonpre}), and it is 0.68 $\mu\Omega$m for LSCO and 1.48 $\mu\Omega$m for OP Bi-2212. Since there is
residual pinning effect indicated by melting field $H_m=1.8$ T in OD LSCO, we assume that some vortices with a density
$H_m/\phi_0$ are pinned and can be simply subtracted, which leads to a substitution of $B+B_T$ with $B-H_m$ in
Eq. (\ref{rhoBpre}).

As shown in Fig. \ref{rhoB}, both the low-field steep rise and high-field rapid saturation of $\rho$ are well captured by Eq. (\ref{rhoBpre}).
{In OD LSCO, since $\eta_0=B_0\phi_0c_0/\rho_n$, which is estimated to be only 1/38 of the BS
model (taking $H_{c2}\approx46$ T) \cite{Wang2006}, the fast vortices scenario is verified at low fields.  The comparison also indicates that $\rho$ saturates at high fields where $\eta$ increases linearly and where the fast vortices scenario breaks down.}
Besides, data (circles) in pseudogap state of OP Bi-2212 \cite{Ri1994} show two behaviors of field dependence of $\rho$ corresponding to the dilute and dense thermal vortices limit. In the dilute thermal vortex limit, $\rho\approx\rho_n B/B_0$, the signal increases linearly with a large slope, from which $B_0$ can be determined. After $B_0$ is measured with the
low-field signal, $B_T$ can be determined from the zero-field signal with Eq. (\ref{rhoBpre}). Near $T_c$
(where $B_T=0$), $\rho$ is determined by magnetic vortices only, thus sharply rises at low field. At $T=105$ K ($>T_c\approx 85$ K, with $B_T\approx45$ T), thermal vortices become dense and $B_T\gg B_0$, which leads to a saturation of $\rho$, thus the sample
behaves metallic-like. This yields a complete picture of the vortex fluid in pseudogap state of HTSC, giving rise to the unconventional field dependence of $\rho$ by Eq. (\ref{rhoBpre}), unifying both the fast vortices scenario and the vortex tangles with core-core collisions. In other words, the magnetoresistance signal in the phase fluctuations regime can be completely described by the current model of vortex tangles; no extra element is needed.

\begin{figure}
\centering
\includegraphics[width=8cm]{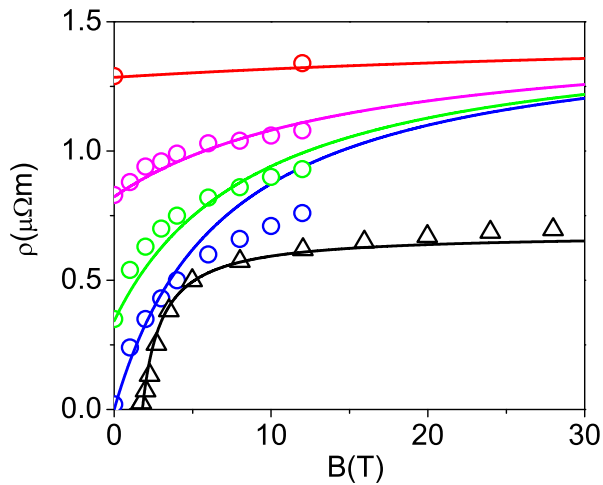}
\caption{The field dependence of $\rho$. Circles are data of OP Bi-2212 at $T=$85 K (blue), 90 K (green),
95 K (magenta), 105 K (red) \cite{Ri1994}. The black triangles are data of OD (p=0.2) LSCO at $T=22$ K ($T_c$=27 K) \cite{Wang2006}.
Solid lines are predictions of Eq. (\ref{rhoBpre}). }
\label{rhoB}
\end{figure}

%\section{Conclusion and Discussion}

The damping model of vortex tangles, Eq. (\ref{etac}) and (\ref{eta}), can be applied to describe other anomalous transport
phenomena in HTSC due to vortex motions. In the absence of a sound damping model in previous studies of Nernst signal in HTSC, only the transverse thermoelectric coefficient $\alpha_{xy}=e_N/\rho$  can be described \cite{Ussishkin2002}. The current damping model enables one to predict the Nernst signal quantitatively if a model of transport entropy is introduced. Using Anderson's proposal
 \cite{Anderson06} and neglecting the pinning effect, {we obtain \cite{Li2017}
\begin{eqnarray}
e_N=C {\rho_nn_s\over T} {B\ln{(H_{c2}/B)}\over{B_0+B+B_T}}.
\label{eNpre}
\end{eqnarray}
where $C=\pi\hbar^2/8m_e\phi_0c_0$, $n_s$ is the two-dimensional superfluid density on Cu-O plane and can be
estimated from a linear model $n_s=n_{s0}(1-T/T_v)$ where $n_{s0}$ is the superfluid density at zero temperature, $T_v$
is the onset temperature of vortex Nernst signal. Near and below $T_c$, $H_{c2}$ is approximately constant, and
$B_T=H_{c2}\exp[-2b(T/T_c-1)^{-1/2}]$ as predicted by Korsterlitz \cite{Kosterlitz1974}. As shown in Fig. \ref{eNB}, the magnitude of the Nernst signals in OD Bi-2212 \cite{Wang2003} are quantitatively described by Eq. (\ref{eNpre}) for a range of
$B$ and $T$. In addition, Eq. (\ref{eNpre}) predicts correctly the peak shift from low to high fields when temperature increases, which is
generated by the increase of thermal vortex density (by $B_T$ in Eq.(\ref{eNpre})). On the
other hard, using the damping model of Anderson or the measurements based on fast vortices
scenario \cite{Anderson06,Bilbro2011}, the prediction will be two orders of magnitude higher than the data at high fields. We conclude that
Nernst signal is also controlled by a damping increase from low to high fields due to vortex tangles, for which our damping model, i.e., Eq. (\ref{eta}), is suitable for quantitative descriptions.}

\begin{figure}
  \centering
  \includegraphics[width=8cm]{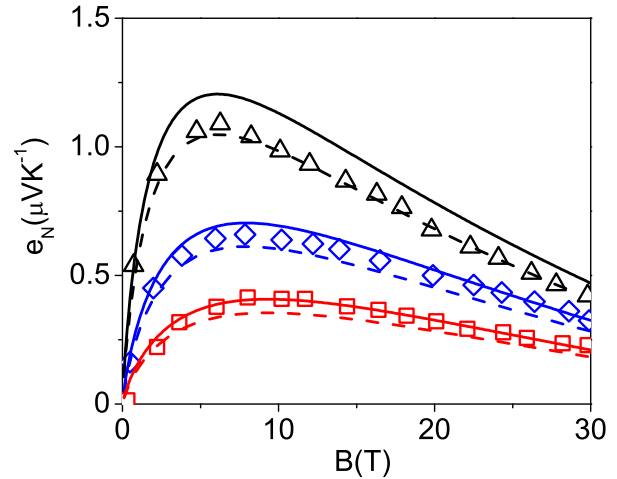}\\
  \caption{$e_N$ vs $B$ at three different temperatures $T_c$=65 K (black), 67.5 K (blue), 70 K (red)). Symbols are data in OD (p=0.22) Bi-2212 \cite{Wang2003}. Solid lines and dash lines are predictions of Eq. (\ref{eNpre}) at $n_{s0}=n_h/2$ and $n_h/2.3$, respectively. The other parameters are $T_v=77.3$ K {(estimated from the onset temperature of $e_N$)}, $\rho_n=0.803$ $\mu\Omega$m (predicted with Eq. (\ref{rhonpre}), see also Fig. \ref{rhon}), and $H_{c2}=50$ T (estimated from zero point of $e_N$),  $B_0=5.5$ T (determined from low
  field signal at $T=65$ K) and $b$=0.25.}
  \label{eNB}
\end{figure}

In summary, {Anderson's vortex tangles concept and phase fluctuation explanation of psedugap are verified in a quantitative manner.}
We go beyond the picture of isolated vortices \cite{Bardeen1965,Halperin1979} with a
novel model of vortex tangles, predicting both flux-flow resistivity and Nernst signal
in cuprate superconductors. Since vortex damping mechanism is the key factor of vortex transport, the current model opens
several new avenues for further studies. First, the model can be extended to describe Fe-based HTSC due to a similar dirty metal
nature. Secondly, distinct from the static calculation with Bogoliubov-de Gennes (BdG)
equations \cite{Canel1965,Melnikov2006}, the new scenario of core-core collision is a
dynamic mechanism, which has implications for the further development of microscopic theories, such as the collision enhancement of qp DOS with a model of vortex-distance fluctuations in qp tunneling calculation. Thirdly, this work suggests that a realistic simulation of vortex fluid at arbitrary $T$ and $B$ by a time-dependent Ginzburg-Landau (TDGL) equation \cite{Machida1993,Kwok2016,Mukerjee2004}
is feasible, if a field and temperature dependent relaxation time is introduced to capture the
dissipation of vortex tangles. Finally, vortex entanglement may yield exotic transport phenomena such as heat transfer in Ettingshausen effects \cite{Palstra1990} and anomalous thermal conductivity in HTSC \cite{Gris2014}.

\end{document}